\documentclass[mathleft,
]{an}
\usepackage{graphicx}
\usepackage{todonotes}
\usepackage{times}
\usepackage{natbib}
\bibpunct{(}{)}{;}{a}{}{,}

\usepackage{amssymb}
\usepackage{xspace}

\newcommand{\chandra}{{\it Chandra}\xspace}
\newcommand{\rxte}{{\it RXTE}\xspace}

\newcommand{\rosat}{{\it ROSAT}\xspace}
\newcommand{\einstein}{{\it EINSTEIN}\xspace}

\newcommand{\suz}{{\it Suzaku}\xspace}
\newcommand{\xmm}{{\it XMM}\xspace}

\newcommand{\astroh}{{\it Astro-H}\xspace}
\newcommand{\erosita}{{\it e-Rosita}\xspace}
\newcommand{\athena}{{\it Athena}\xspace}

\newcommand{\swift}{{\it Swift}\xspace}
\newcommand{\INTE}{\textit{INTEGRAL}\xspace}
\newcommand{\nus}{\textit{NuSTAR}\xspace}

\newcommand{\msun}{\ensuremath{M_{\odot}}\xspace}

\newcommand{\lx}{\ensuremath{L_{\rm X}}\xspace}

\newcommand{\ledd}{\ensuremath{L_{\rm Edd}}\xspace}

\newcommand{\ergs}{\ensuremath{{\rm ergs~s}^{-1}}\xspace}
\newcommand{\Mone}{M82\,X-1\xspace}
\newcommand{\Mtwo}{M82\,X-2\xspace}
\newcommand{\Moneother}{M82\,X41.4+60\xspace}
\newcommand{\Mtwoother}{M82\,X42.3+59\xspace}

\newcommand{\grs}{GRS\,1915+105\xspace}
\newcommand{\asec}{\ensuremath{^{\prime\prime}}\xspace}

\newcommand{\nat}{Nature\xspace}
\begin{document}

\Pagespan{001}{}
\Yearpublication{201X}%
\Yearsubmission{2015}%
\Month{XX}%
\Volume{XXX}%
\Issue{XX}%
\DOI{This.is/not.aDOI}%

\title{Ultraluminous X-ray sources: three exciting years}

\author{M. Bachetti\thanks{\email{bachetti@oa-cagliari.inaf.it}\newline}
}
\titlerunning{ULXs: three exciting years}
\authorrunning{M. Bachetti}
\institute{
INAF/Osservatorio Astronomico di Cagliari, via della Scienza 5, I-09047 Selargius (CA), Italy}

\received{XX XXX 2015}
\accepted{XX XXX 201X}
\publonline{later}


\abstract{%
The extreme extragalactic sources known as Ultraluminous X-ray Sources (ULX) represent a unique testing environment for compact objects population studies and the accretion process.
Their nature has long been disputed. Their luminosity, well above the Eddington luminosity for a stellar-mass black hole, can imply the presence of an intermediate-mass black hole or a stellar black hole accreting above the Eddington limit.
Both these interpretations are important to understand better the accretion process and the evolution of massive black holes.
The last few years have seen a dramatic improvement of our knowledge of these sources.
In particular, the super-Eddington interpretation for the bulk of the ULX population has gained a strong consensus.
Nonetheless, exceptions to this general trend do exist, and in particular one ULX was shown to be a neutron star, and another was shown to be a very likely IMBH candidate.
In this paper, I will review the progress done in the last few years.
}

\maketitle

\section{Introduction}
Ultraluminous X-ray sources are extragalactic point-like, off-nuclear X-ray sources whose apparent X-ray luminosity \lx exceeds the Eddington limit for a 10\msun black hole (BH), or $\lx\gtrsim10^{39}\ergs$.
The first evidence for these sources came from the \einstein observatory \citep{Long:1983}.
After a number of different surveys we nowadays know some hundreds of them (e.g., from \chandra, \citealt{Swartz+04}; \rosat, \citealt{LiuBregman05}; \xmm, \citealt{Walton+11}).
Two main classes of models have naturally arisen from the observation of these very luminous sources.

The first class of models involves black holes of larger mass than stellar remnant BHs, accreting in the same sub-Eddington regime as the well-known Galactic BHs \citep[etc.]{Kaaret+01, Miller+03}.
This means that ULXs would be rare examples of intermediate-mass black holes (IMBHs), with masses $\gtrsim100$\,\msun.
These objects are more massive than expected from a single star collapse \citep[e.g.][]{Belczynski+10}, and possible mechanisms for their formations include the runaway collapse of a cluster of stars \citep{PortegiesMcMillian02}, and remnants of primordial stars \citep[e.g.][]{MadauRees01,brommlarson04}.
Evaluating the number of IMBHs has profound implications for the models of the evolution of supermassive black holes (SMBH).
In fact, a possible path for the growth of SMBHs is through the merger of smaller, ``seed'' BHs, represented indeed by IMBHs \citep{KormendyHo13}.
This class of models is very likely to describe the most extreme of these sources, called hyperluminoux X-ray sources.
The most famous of this kind is ESO 243-49 HLX-1 \citep{Farrell+09}, with a luminosity above $10^{42}$\,\ergs, whose behavior is also consistent with part of the phenomenology of standard black holes.
For example, it undergoes spectral transitions and outbursts, similar to those of Galactic black holes, with spectral states similar to this standard picture.
Jets are also observed during spectral transitions.
Only the time scale of these outbursts does not fit in this simple scenario (recurrence timescale: \citealt{Lasota+11}; irregularity of timescale: \citealt{Godet+14}).

The second class of models involves stellar black holes, with a super-Eddington apparent luminosity.
Real super-Eddington accretion might be achieved up to 10\,\ledd through the so-called photon-bubble instability in standard ``thin'' disks \citep{Begelman02} or in inefficient regimes of accretion like the ``slim'' disk \citep[e.g.][]{Kawaguchi03}.
Mild beaming \citep[e.g.][]{King+01} might account for another factor and permit luminosities up to $10^{41}$\,\ergs without requiring IMBHs.
Models involving extreme beaming, for example the fact of looking inside a jet \citep[e.g.][]{Koerding+01}, are mostly ruled out from the observation of fairly isotropic optical bubbles (see Section \ref{sec:outflows}).
This class of models, putting forward the hypothesis of extreme mass transfer, is also very interesting.
In fact, it influences the timescales of the evolution of black holes from the initial seeds to the SMBHs we observe today \citep{Kawaguchi+04,ReesVolonteri07,Volonteri:2010ik}.

Therefore, ULXs might be an important piece of the cosmological puzzle, permitting a test study of two phenomena relevant to black hole evolution: IMBHs and super-Eddington accretion.

The definition of ULXs is based only on one observable (the apparent isotropic luminosity), so it is also likely that these objects are not truly a class, but more like a``zoo'' with different animals.
The findings of the last three years seem to confirm this, as we are going to see.

Since we know some hundred ULXs, it is not surprising that many subclasses have appeared in the literature, mostly based on luminosity ranges.
Since there is some level of inconsistency between the definitions given to these subgroups of ULXs in different papers, in this work we will consider {\em weak} ULXs (wULX) those radiating below $10^{40}\ergs$, {\em strong} ULXs (sULX) those above that and below $10^{41}\ergs$, {\em extreme} ULXs (eULX) those below $10^{42}\ergs$, and {hyperluminous} X-ray sources (HLX) those above $10^{42}\ergs$.
As anticipated, this review will cover mostly the range from wULXs up to eULX, where the overlap between the IMBH and the super-Eddington interpretations is larger.

The history and general properties of ULXs can be found in several other reviews (e.g. \citealt{Fabbiano06}; \citealt{FengSoria}; \citealt{Webb:2014vj}; \citealt{Bachetti+15rev} in prep.).
In this review, I will concentrate on the discoveries and progress in the understanding of these objects gained in recent times (three/four years).

\begin{figure*}[htb]
\centering
\includegraphics[width=2.3in]{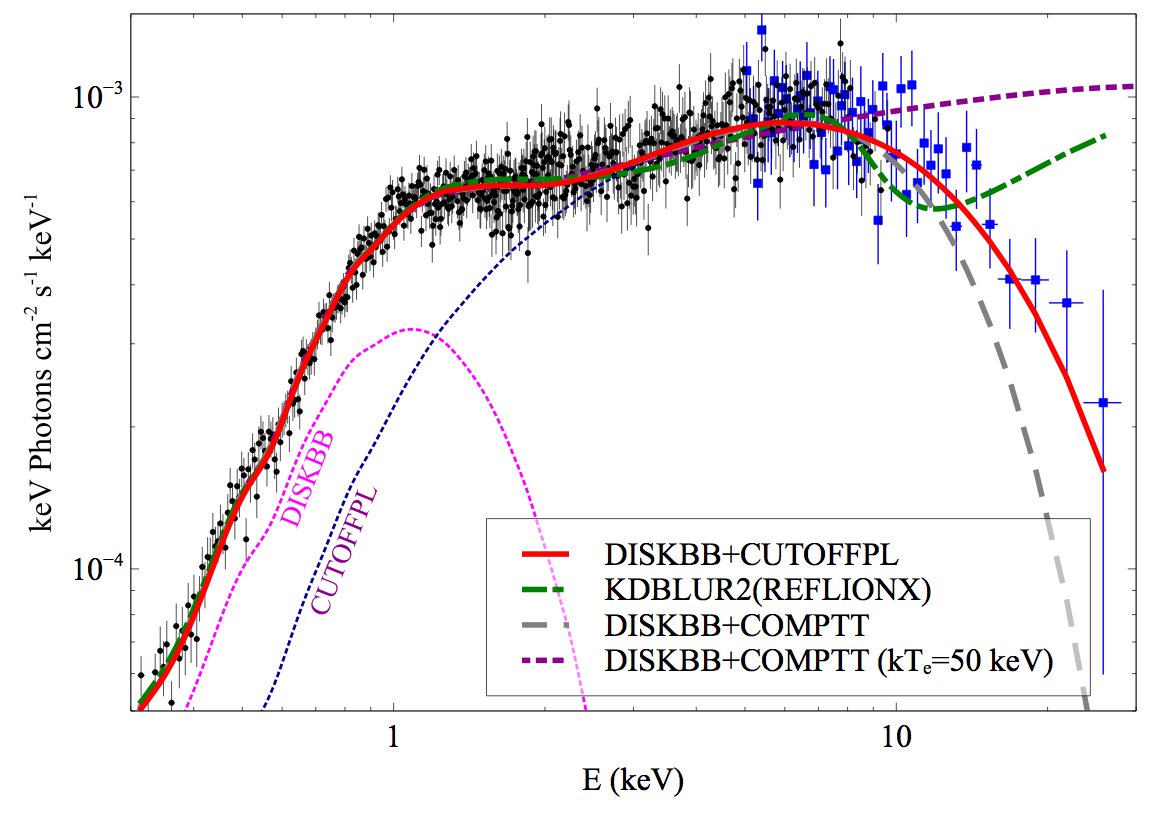}
\includegraphics[width=1.9in]{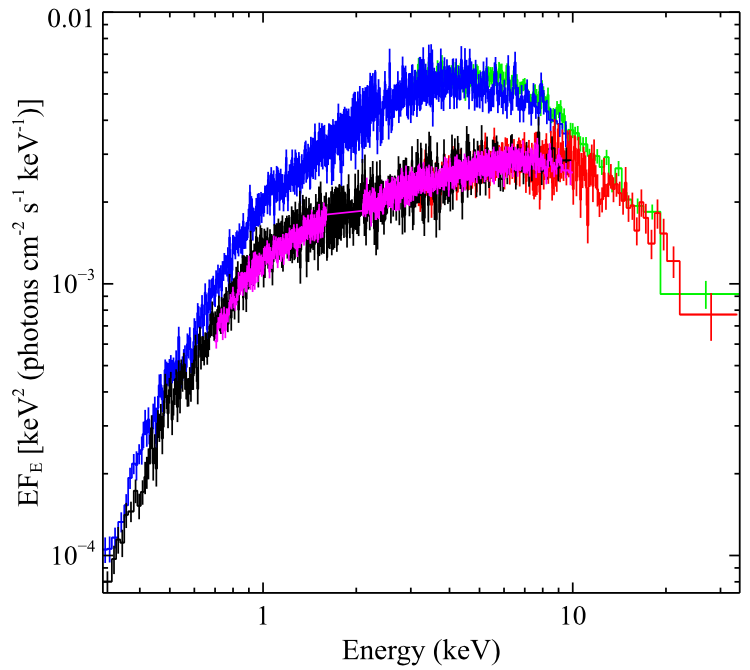}
\includegraphics[width=2.3in]{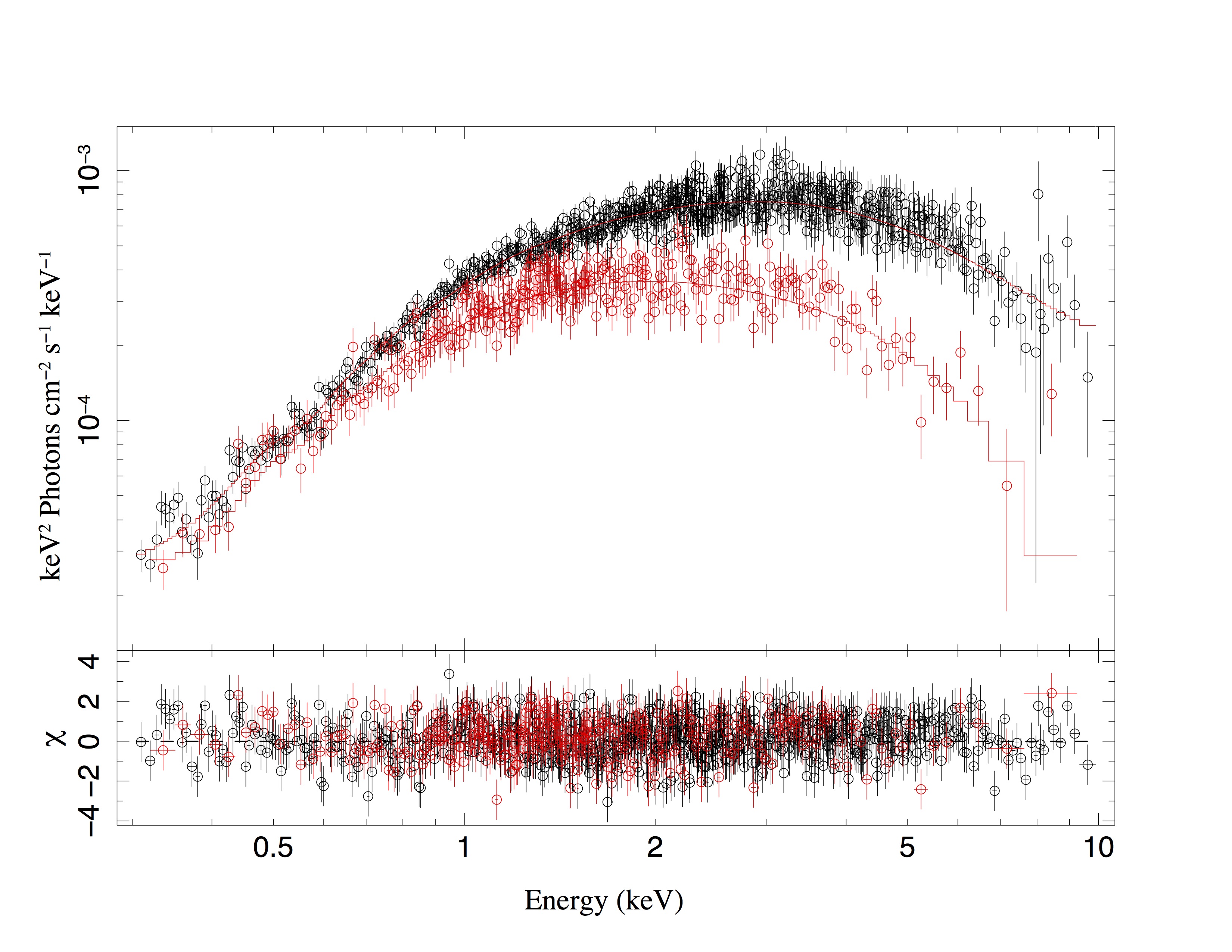}
\caption{A sample of NuSTAR ULXs, showing the ubiquitous curved spectrum at high energy and the strong spectral variability observed in several sources of the sample (from \citealt{Bachetti+13} and \citealt{Walton+14}).
(Left) \xmm and \nus spectra of NGC 1313 X-1, showing the typical hard ultraluminous shape, the cutoff ruling out power law and reflection, and the excess with respect to single-T comptonization. (Middle and right) Holmberg IX X-1 and NGC 1313 X-2, showing the extreme variability observed in several sources on short time scales.}
\label{fig:nus}
\end{figure*}

\section{Ultraluminous State?}\label{sec:states}

In ULXs, the spectral and timing properties seem to be consistent with three main ``states'' \citep{Sutton+13}: a broadened disk state, with a single thermal component at some keV, and two so-called ultraluminous states, containing a low-energy soft excess (0.1--0.3\,keV) and a power law-like component with a slight downturn above 5 keV.
These two ultraluminous states are named soft and hard ultraluminous, and they differ only on the slope of the power law.
If the excess and the power law tail are to be interpreted as standard black hole spectra, the well known inverse proportionality between disk temperature and mass \citep{SS73} would suggest that these spectra are indeed from a IMBH \citep[e.g.][]{Miller+03,Miller+04}.
The observation of standard transitions from a disk-dominated to power-law dominated state would point towards this interpretation.
However, even if strong luminosity variations are known in ULXs, transitions between dramatically different spectral {\em shapes} are very rare \citep[e.g.][unsurprisingly a strong IMBH candidate]{FengKaaret10}.
A few more have been shown to transition between states classified as ultraluminous \citep{Sutton+13}, but as the sampling of the ULX population gets better, more are found \citep{Walton+13,Walton+14}.
Also, the broadened disk is more likely to be observed in wULXs, and the two sources (NGC 1313 X-1 and Holmberg IX X-1) showing a transition between hard and soft ultraluminous, the soft was at higher fluxes than the hard. Fast variability is usually observed only in the soft ultraluminous and broadened disk states \citep{Sutton+13}.

In this super-critical accretion scenario, interpretations of the soft and hard components of the spectrum are very different from standard BH spectra: in the \citet{Gladstone+09} interpretation, the disk was invisible, the hard curved component was Comptonization of the underlying disk from an optically thick corona, while the soft component, more prominent at high luminosities, was produced by the far away truncated disk (outside the corona) and by winds arising at the extreme accretion rates.
According to a more recent intepretation \citep{Sutton+13}, instead,  the hard component is related to the temperature of the inner disk, while the soft component is arising from the wind, that reprocesses the disk emission and partially occultates it.
In this interpretation, the hard component is variable because the clumpy nature of the wind occultates randomly the inner region, imprinting variability on an otherwise more or less stable flux.
This interpretation is now gaining consensus, in particular because it can explain both timing and spectral properties of ULXs \citep[e.g.]{Middleton+15}.
The expected signatures of outflows in ULXs have been elusive. See \citep{Roberts15} for recent developments, and Section\,\ref{sec:outflows} for more details.

\section{Weak ULXs - proofs of super-Eddington accretion}\label{sec:llulx}

ULXs exceeding by less than an order of magnitude the Eddington limit were the easiest to explain with slightly larger black holes or slightly above-Eddington accretion.
Nonetheless, these sources are numerous and they have been protagonists of some of the most important developments in the last few years.

One such example is the source XMMU J004243.6+412519 in M31. Discovered by \citet{2012ATel.3890....1H} at \lx$\gtrsim10\cdot10^{38}$\,\ergs, it showed an increase in luminosity up to ULX levels in two subsequent detections.
\citet{Middleton+13} performed a joint X-ray/radio monitoring with \xmm and the VLA.
They found highly variable radio emission on timescales of tens of minutes, implying a very compact source ($\lesssim5$\,AU).
Also, whereas the spectrum could be fit with models implying either standard accretion disks or ULX broadened disks, the {\em behaviour} was not consistent with a standard $L\propto T^4$ relation expected from the standard disk.
The comparison of these properties with known Galactic X-ray binaries such as \grs, lead to the identification of this source as a StBH undergoing a transition to the super-Eddington regime.

\citet{Liu+13} found an optical modulation due to orbital motion, with period 8.2\,d, of M101 X-1, a ULX radiating at $\approx 3\cdot 10^{39}\ergs$.
Together with the observation that the companion is a Wolf-Rayet star, the estimated mass range is $5<M<20$\msun.
The authors find signatures that accretion is happening from a stellar wind rather than Roche-Lobe overflow.
\citet{Rong-Feng+15} find for this source signatures of a thick outflow.

Finally, \citet{Motch+14} found that the source P13 in NGC 7793, showing all typical spectral signatures of ULXs (curved spectrum, soft excess, \lx$\approx 4\cdot 10^{39}\ergs$), really is a black hole with mass $< 15$\msun.
This was done through the measurement from optical observations of the orbital period of 64\,d, together with the identification of the companion star as a B9Ia star.
This is considered some of the best evidence that the curved spectra of ULXs (see Section~\ref{sec:nus}) are a signature of supercritical accretion.

\section{NuSTAR: yes, it's a cutoff}\label{sec:nus}

One of the main questions about ULX spectra, before 2011, was whether the downturn above 5\,keV \citep{Stobbart+06} was a real cutoff, produced through Comptonization from a cold, optically thick corona \citep{Gladstone+09} or the effect, for example, of a broadened iron line (e.g. \citealt{CaballeroGarcia+10}) over a power law continuum.
The first hypothesis pointed strongly towards a new accretion regime, probably related to super-Eddington accretion, while the second was a possible way to justify the downturn in the IMBH scenario.

However, the spectral coverage granted by \xmm, \chandra, \swift, limited to 10\,keV, was not sufficient to disentangle between these very different models \citep{Walton+11}.
Non-imaging Hard X-ray satellites like \INTE or \suz (as was done later by \citealt{Yoshida+13,Dewangan+13,Sazonov+13}),
forced to rely on very heavy assumptions and very uncertain background subtraction procedures.
For extragalactic sources like ULXs one can rarely assume that the target dominates the emission over the field of view (as one would do for most Galactic sources outside some well-determined dense regions).
From this point of view, the launch of \nus was a breakthrough in ULX studies.
Its imaging capabilities and spectral coverage up to 79\,keV, with a comparable effective area to \xmm in the 5--10\,keV range, permitted to run a series of large programs of \nus observations, aided by the soft X-ray coverage of \xmm, \swift or \suz, obtaining the first broadband (from 0.3 to 40keV) X-ray spectra of these objects and measuring the spectral and timing variability when present.
This program was able to clearly show that a real cutoff was present in all sULXs and eULXs of the program (e.g. \citealt{Bachetti+13}, \citealt{Walton+13}, \citealt{Rana+15}, \citealt{Walton+14}; this favored an interpretation of these ULXs as StBHs (probably in the high mass range for this class) accreting around or above the Eddington limit.
Moreover, in most of these works the cutoff was found in excess of the prediction from Comptonization by a single-temperature corona, that was hypothesized in some papers \citep[e.g.][]{Gladstone+09}.

\section{M82 - a cradle of exceptions}
\begin{figure*}[t]
\centering
\includegraphics[width=3in]{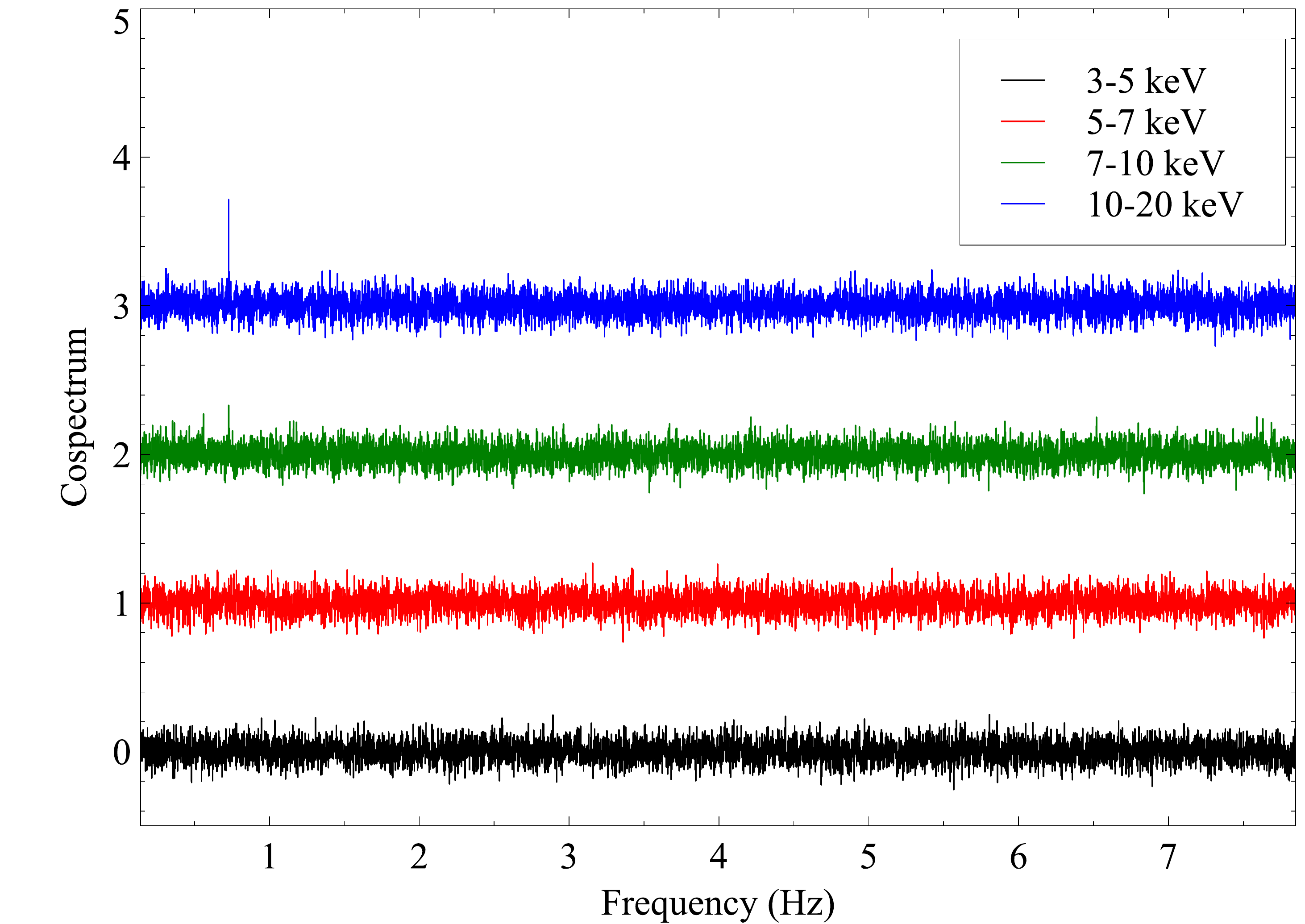}
\includegraphics[width=3.in]{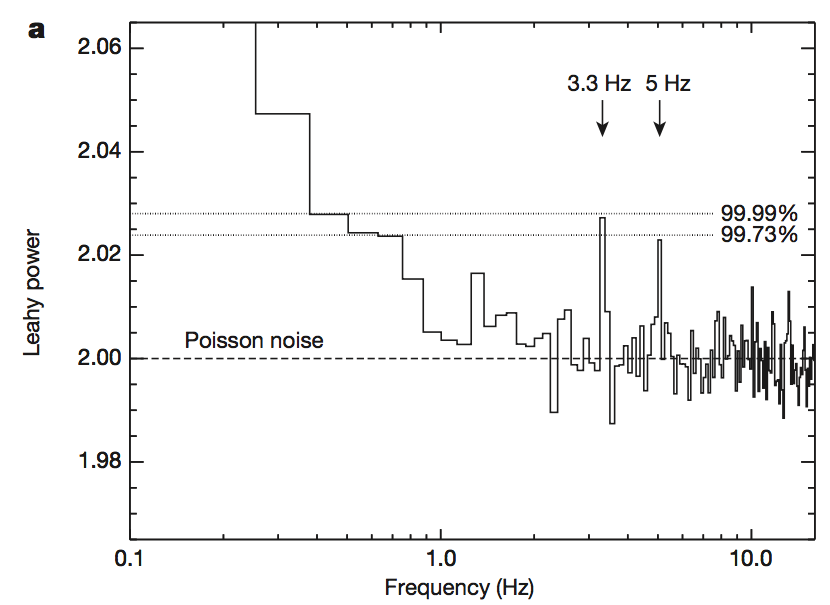}
\caption{(Left) 
The cospectrum \citep[\'a la][]{Bachetti+15} of one of the long obsIDs of the \nus campaign of M82, showing the peak at $\sim0.7$\,Hz corresponding to the pulsar, in different energy bands.
(Right) \rxte power spectrum from \citet{Pasham+14} showing the pair of high-frequency QPOs from \Mone. The 5\,Hz QPO is above detection level with a different binning.
}
\label{fig:m82}
\end{figure*}

In 2014, two remarkable discoveries came out of the Cigar Galaxy, M82, that harbors three known ULXs \citep{Matsumoto:2001fu, Kaaret+01, FengKaaret07, Kong+07, jin+10}.
The first was the observation of quasi-periodic oscillations from a known eULX, \Mone, also known as \Moneother%
\footnote{Sources in M82 are often named by their offset from $\alpha=09^h51^m00^s$, $\delta=+69\deg54\arcmin00\arcsec$ (B1950.0)}%
. The second was the discovery of pulsations from a sULX just 5\asec away from \Mone, \Mtwo (or \Mtwoother).

\Mone is a very well-known variable eULX, reaching above $10^{40}$\,\ergs \citep{PtakGriffiths99,Kaaret+01}.
It was observed to undergo spectral transitions reminiscent of standard BH spectral states (transition to ``thermal-dominant``: \citealt{FengKaaret10}), and this pointed strongly towards the IMBH interpretation.
It's one of the few ULXs known to show strong quasi-periodic oscillations, detected by \rxte and \xmm in the range $50-100$\,Hz \citep{Mucciarelli+06}.
In 2014, the IMBH hypothesis gained strong support when a timing analysis including all \rxte observations of \Mone showed a new pair of quasi-periodic oscillations, at $\sim 3$ and $\sim 5$\,Hz \citep[see Figure\,\ref{fig:m82}]{Pasham+14}.
The frequencies of these oscillations were consistent with a 3:2 ratio observed in the high-frequency QPOs of two Galactic Black holes at hundreds of Hz \citep[but whose identification is unclear, as is the scaling with the mass, see ][]{Belloni+12}.
If this identification is correct, a simple scaling of the frequencies leads to a mass estimate of $\sim$400 \msun.
This makes \Mone one of the strongest IMBH candidate in the eULX range.

But the most unexpected result was probably the discovery of the first ULX powered by an accreting neutron star \citep[see Figure\,\ref{fig:m82}]{Bachetti+14}.
This source was a well-known ULX, showing very strong luminosity variations on timescales of $\sim$weeks and up to $3\cdot10^{40}$\ergs \citep{Kong+07,FengKaaret07}.
The presence of mHz QPOs had been used to model it as an IMBH above 10000\msun \citep{Feng+10}.
Pulsations, unequivocally, identified it as a NS.
The possible explanations for the extreme luminosity of this object, 100 times the Eddington limit for a neutron star and $\sim$10 times higher than the limiting luminosity for a NS \citep{BaskoSunyaev76}, include the changes in the Thomson scattering coming from a strong magnetic field (e.g. \citealt{Eksi+15, Dallosso+15}), and beaming (e.g. \citealt{Christodolou+14}).
This source has also been proposed as an alternative path for the formation of millisecond pulsars \citep{KluzniakLasota15}.

\section{Jets and Outflows}\label{sec:outflows}
\begin{figure}[t]
\centering
\includegraphics[width=3in]{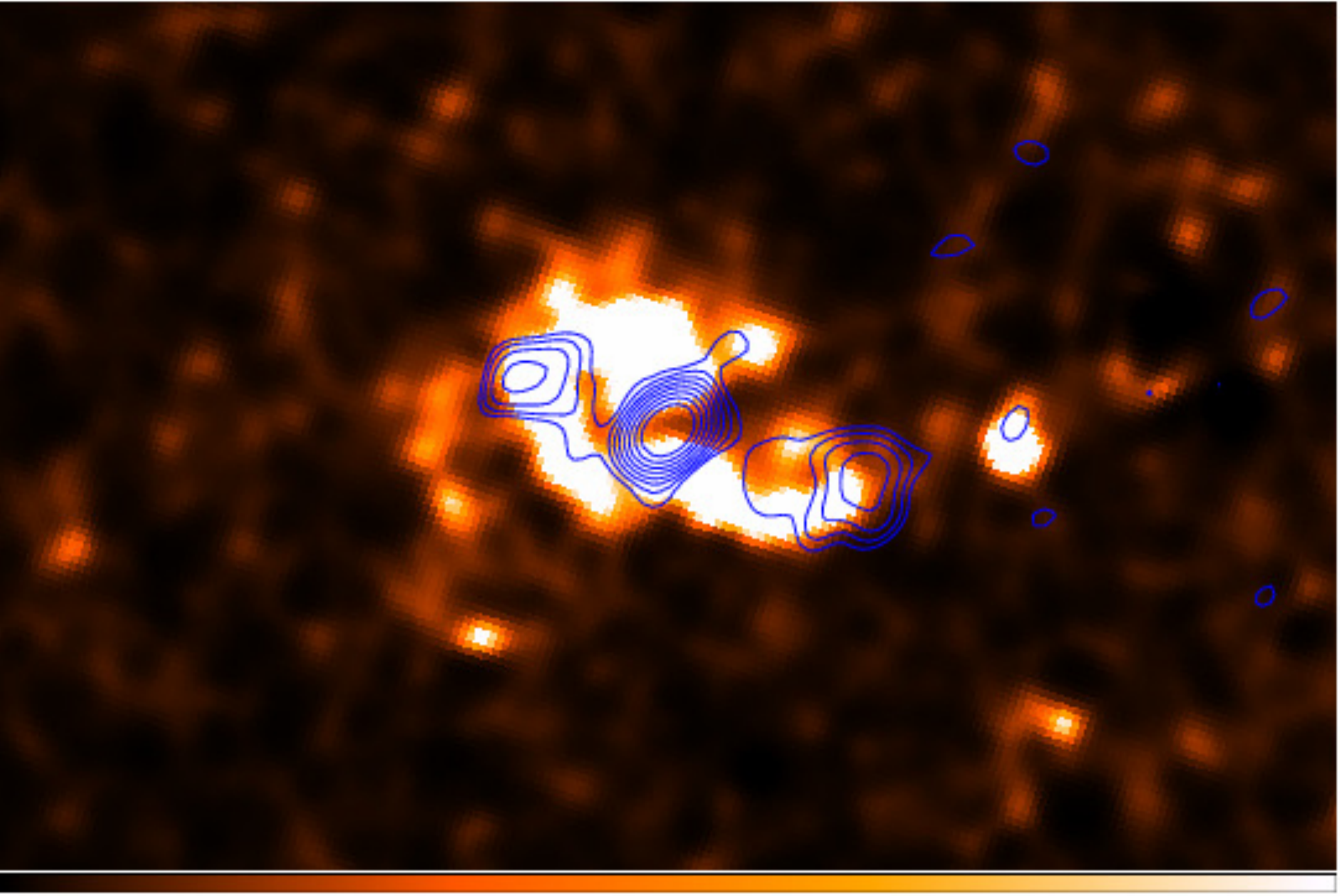}
\caption{HST image using the FR462N narrow-band filter of the optical bubble around Holmberg II X-1, with contours showing the radio structure associated with a jet \citep{Cseh+14}.}
\label{fig:bubbles}
\end{figure}
Collimated jets are usually observed in hard and intermediate states of sub-Eddington Galactic BHs, but not in the high/soft states \citep[]{Fender+04}.
In the microquasar \grs, often compared to ULXs for its peak luminosity, steady jets are present during the hard so-called ``plateau'' state and ejection events during state changes \citep{1994Natur.371...46M,2004ARA&A..42..317F}.
Jets are  associated with HLXs during state transitions, and this is considered evidence of the sub-Eddington regime, and so of their nature as IMBH \citep{Cseh+15hlx1,Mezcua+15}.
Only one sULXs, Holmberg II X-1, has been shown to have jets in ultraluminous states \citep{Cseh+14, Cseh+15}.
The mechanism of formation of jets is still not well understood.
Simulations do confirm that, in some setups, jets appear in super-Eddington accreting objects \citep[e.g.][]{2015arXiv150802433M}.

Strong outflows are instead generally expected to arise from super-critical accretion (e.g., in simulations: RHD, \citealt{OhsugaMineshige11, Hashizume+15}; GRMHD, \citealt{McKinney+14}), and the presence of these outflows, cold and/or optically thick, have been proposed alternatively as the origin for the soft excess in ULX spectra \citep[e.g.][]{King+01, Begelman02, KingPounds03, Kubota+04, Gladstone+09, FengSoria}.

As we've seen, in the \citet{Sutton+13} framework it is expected that these outflows play an important role both in the spectral and timing behavior of ULXs (see Section\,\ref{sec:states}).

However, direct observations of outflows have proven, to-date, elusive.
Until now, most of the evidence of outflows came from shock-ionized and X-ray ionized nebulae, seen in optical and radio observations and ruling out anisotropic accretion for most ULXs \citep[e.g.]{PakullMirioni02,2003Sci...299..365K,Lehmann:2005gx,Kaaret:2009ky}.

Signatures in the Fe K complex often observed in ultra-fast outflows from SMBHs \citep[e.g.][]{Tombesi+10} and BH and NS binaries \citep[e.g.][and references therein]{2012MNRAS.422L..11P} are not found in ULXs \citep{Walton+12}.
\citet{Roberts15} investigate excesses often seen in X-ray spectra \citep[e.g.][]{2009ApJ...703.1386S,Miller+13,Bachetti+13} as possible signatures of outflows.

\citet{Fabrika+15} report on optical observations of ULXs, where they show that the spectra of ULXs, very similar to each other, originate from very hot winds from the accretion disks.
The optical spectra are indeed similar to that of the Galactic source SS 433 but with a higher wind temperature.
This points towards the supercritical stellar black holes interpretation.

\section{Conclusions}
\citet{FengSoria}'s famous review on ULXs in 2011 more or less predicted correctly the landscape we now know of ULXs:
\begin{quote}
{\em ULXs are a diverse population; MsBHs with moder-
ate super-Eddington accretion seem to be the easiest
solution to account for most sources up to luminosi-
ties $\sim$a few $10^{40}$\,\ergs; strong beaming ($1/b>10$) can be ruled out for the majority of ULXs; IMBHs are preferred in a few exceptional cases}
\end{quote}

Nonetheless, reading in detail that review there are observing properties that changed, several open questions that have been addressed, and interpretations that evolved:
\begin{itemize}
\item ULX spectra {\em do} vary significantly with flux increases. More and more ULXs were found to change their spectral shape considerably and some of them even to undergo dramatic luminosity increases on timescales of $\sim$weeks \citep{Bachetti+13,Walton+14,Walton+15}.
\item Neutron stars were only mentioned twice, in the same phrase, and not as possible ULX-powering compact objects. The discovery of \Mtwo was completely unpredicted.
\item The toy model about ULX emission gave the soft emission coming from the outer disk and the hard emission from the inner disk and the wind. Today's leading interpretation interprets the soft component as coming from the wind.
\end{itemize}

Also, some of the bullet points of possible evolution of ULX studies have been addressed, in particular (letters are referred to the original article points, and italic is used for the original text):
\begin{itemize}
\item[(a)] {\em (...) search for possible high frequency fea-
tures (breaks and QPOs) that are found in Galactic BHs at
frequencies $\sim10^2$\,Hz}: \Mone was indeed interpreted as an IMBH thanks to the discovery of QPOs in a 3:2 ratio
\item[(b)] {\em Determining  the  relative  contribution  of  thermal  emission and Comptonization component is a key test (...) X-ray  telescopes  with  good  sensitivity  up  to  a  few  tens  of  keV  are needed}: \nus proved to be capable of doing this, clearly finding an excess of the cutoff from the predictions of single-T comptonization.
\item[(i)] {\em searching for compact radio jets (...)}: compact radio jets were indeed found in a eULX \citep{Cseh+14,Cseh+15} and in a HLX \citep{Mezcua+15}.
\end{itemize}

For the remaining questions in the \citet{FengSoria} review, the landscape of the next few years looks encouraging.
The launch of \astroh with its high spectral resolution and good imaging capabilities \citep{astroh14stmbh}, the surveys by \erosita \citep{2012arXiv1209.3114M} approaching, SKA \citep{Wolter+15ska} in the works, and the next big X-ray observatory, \athena \citep{Nandra+13} to come in the 2028, will surely fill most of the instrumental gaps that slowed down the progress until now.

As it often happens, the discoveries have opened the path to new questions and new important fields of investigation.
Super-Eddington accretion is now accepted as a relatively frequent phenomenon.
There is much to be learned about it yet: how frequent it is, how it changes the accretion geometry, if it changes considerably the evolution time scales of black holes and galaxies.

Timing techniques will be likely to gain importance. Besides being key for the two major discoveries in M82, they represent an independent and complementary approach to spectral studies. Spectral timing studies of ULXs, for example based on time lags \citep[e.g.]{DeMarco:2013vp} and covariance spectra \citep[e.g.]{Middleton+15} are very promising.
A thorough search of pulsations in ULXs is already ongoing from several groups \citep[e.g.][]{Doroshenko:2015vb}.
This is not an easy task; ULXs are distant sources, their flux is relatively low, their signal often contaminated, and detection limits are heavily dependent on flux and rms \citep{Lewin+88}.
Nonetheless, it's probable that other neutron stars will be found in ULXs, thanks to the upcoming focusing telescopes and the awareness that this is an option.

\acknowledgements

MB was supported in part by an {\em assegno di ricerca} from the Sardinian Region. The author wishes to thank Fiona Harrison, Dom Walton, Didier Barret, Natalie Webb, Tim Roberts, Andrea Tarchi, Paola Castangia and the \nus ULX working group, for many discussions that posed the bases for this work, and the referee for his/her careful reading and very helpful comments.

\bibliographystyle{meoshort}

\end{document}